\documentclass[doublecol]{epl2} 
\usepackage{amssymb}
\usepackage{amsfonts}
\usepackage{amsmath}
\usepackage{amsthm}
\usepackage{bbm}
\usepackage{xcolor}

\title{Relay synchronization in multiplex networks of discrete maps}
\shorttitle{Relay synchronization in multiplex networks of discrete maps} %Insert here a short version of the title if it exceeds 70 characters

\author{M. Winkler\inst{1} \and J. Sawicki\inst{1} \and  I. Omelchenko\inst{1} \and A. Zakharova\inst{1} \and V. Anishchenko\inst{2} \and E. Sch{\"o}ll\inst{1}}
\shortauthor{M. Winkler \etal}

\institute{                    
  \inst{1}Institut f{\"u}r Theoretische Physik, Technische Universit\"at Berlin, Hardenbergstra\ss{}e 36, 10623 Berlin, Germany\\
  \inst{2}Department of Physics, Saratov State University, Astrakhanskaya street 83, 410012 Saratov, Russia\\
}

\abstract{Complex multiplex networks consist of several subnetwork layers, which interact via pairwise inter-layer connections. 
Relay synchronization between distant layers which are not directly connected, but only via a relay layer, can be observed in multiplex networks. We study three-layer networks of discrete logistic maps, where each individual layer is a nonlocally coupled ring, and demonstrate scenarios of relay synchronization of complex patterns in the outer layers which interact via an intermediate layer. We find regimes of 
relay synchronization for chimera states, i.e., patterns of coexisting coherent and incoherent domains, and a transition from phase chimeras to amplitude chimeras for increasing inter-layer coupling. We determine analytically the approximate critical coupling strengths for the existence of phase chimeras.}

\begin{document}

\maketitle

\section{Introduction}
Complex networks have a wide range of applications in physics, chemistry, biology, neuroscience, technology, and social sciences \cite{STR01a,ALB02a,NEW03,SCH16}. Of particular interest are synchronization phenomena \cite{PIK01,BOC06a,BOC18}, including full synchronization and more complex partial synchronization patterns \cite{PAN15,SCH16b} like clusters, chimera states, or solitary states, as a result of the interplay of the dynamics of individual nodes and topology. Chimera states are intriguing partial synchronization patterns, first discovered in networks of identical phase oscillators with a symmetric coupling function \cite{KUR02a,ABR04}. In spite of the symmetric topology, symmetry-breaking collective behaviour can spontaneously result in patterns combining both spatially coherent and incoherent domains.

Since their initial discovery, chimera states have evoked great interest \cite{MOT10,BOR10,OME10a,OME12a,MAR10,WOL11a,BOU14,SET14,FEN15} and have been found in numerous models, among them time-discrete maps with chaotic and periodic dynamics \cite{OME11,SEM15a,BOG16,BOG16a,VAD16,SEM17,SHE17c,SHE17,BUK17,BUK18a}, time-continuous chaotic systems \cite{OME12,SEM15a}, neural systems \cite{OME13,ROT14,OME15,SEM16,AND16,AND17}, Boolean networks \cite{ROS14a}, population dynamics \cite{HIZ15}, and quantum oscillator systems \cite{BAS15}. Chimeras have been successfully verified in experiments, including optical \cite{HAG12}, chemical \cite{TIN12,NKO13}, mechanical \cite{MAR13,KAP14}, electronic and optoelectronic \cite{LAR13,LAR15} and electrochemical oscillator networks \cite{WIC13,WIC14,SCH14a}. Further, the robustness of chimera states has been demonstrated in networks of inhomogeneous oscillators \cite{LAI10}, or with irregular topologies  \cite{KO08,SHA10,LAI12,YAO13,ZHU14,SCH16b,OME15,HIZ15,ULO16,TSI16,SAW17}. 

Many applications in nature and technology, such as neuronal and genetic networks, transportation networks, power grids, and social networks, require multilevel topologies~\cite{DOM13,KLE16,MAK16,GHO16,DOM17,BAT17,GOR17,MAJ17,LEY17a,GHO18,AND18a,CAR18a}. Recent studies have focused on the analysis of synchronization scenarios in multilayer networks, such as remote and relay synchronization~\cite{NIC13,ZHA17,LEY18,SAW18,SAW18c}, when distant layers which do not have direct connections tend to synchronize. Multiplex networks are a subclass of multilayer networks, having an identical set of nodes in each layer, and allowing for only pairwise inter-layer links between corresponding nodes of the neighbouring layers~\cite{KIV14}. It has been shown recently that multiplexing can be used as an instrument to control spatio-temporal patterns in networks~\cite{GHO16a,JAL17,GHO18,MIK18,OME19}, when the desired state in a certain layer is achieved without direct manipulation of this layer, but with manipulating the neighbouring layers. For instance, even weak multiplexing can induce coherence resonance~\cite{SEM18}, chimeras, and solitary states~\cite{MIK18} in neural networks. 

In this letter, we study relay synchronization in three-layer multiplex networks of time-discrete logistic maps, where the individual uncoupled maps are characterized by chaotic dynamics, and inside each layer we have a nonlocally coupled ring. We demonstrate the synchronization of chimera states in the two outer layers of our network, which interact via the intermediate layer. Moreover, we study the transition from phase to amplitude chimeras with increasing inter-layer coupling strength and provide analytical conditions for phase chimeras in terms of the network coupling strengths.

\section{\label{sec:Model}The Model}

We study a multiplex network consisting of three layers, each represented by a ring of non-locally coupled time-discrete logistic maps:
\begin{eqnarray}\label{eq:multiplex}
    z^m_i(t+1) = &\underbrace{f(z^m_i(t))}_{\text{local dynamics}} \\ 
    + &\underbrace{\frac{\sigma_m}{2R_m} \sum^{i+R_m}_{j=i-R_m}[f(z^m_j(t)]-f(z^m_i(t)]}_{\text{intra-layer coupling}} \nonumber\\[-1mm]
    + &\underbrace{\sum^{3}_{l=1}\sigma_{ml}[f(z^l_i(t))-f(z^m_i(t))]}_{\text{inter-layer coupling}} \nonumber
\end{eqnarray}
\noindent $z^m_i$ are real dynamical variables, with node index $i = 1,...,N$; all indices $i,j$ are modulo $N$. The layer index is given by $m=1,2,3$. The discrete time is denoted by $t$; $f(z)$ is a one-dimensional logistic map with $f(z) = az(1-z)$, where we fix the bifurcation parameter $a = 3.8$, corresponding to chaotic dynamics of the individual uncoupled unit. $R_m$ is the non-local intra-layer coupling range,  associated with the dimensionless coupling radius $r_m = \frac{R_m}{N}$, and $\sigma_m$ is the intra-layer coupling strength. For an ordinal inter-layer coupling with constant row sum we choose the inter-layer coupling matrix: 
\begin{equation}\label{eq:coulpingmatrix}
    \sigma_{ml} = \left( {\begin{array}{ccc}
                  0 & \sigma_{12} & 0\\
                  \frac{\sigma_{12}}{2} & 0 & \frac{\sigma_{32}}{2} \\
                  0 & \sigma_{32} & 0 \\
                    \end{array} } \right)
\end{equation}
with $\sigma_{12} = \sigma_{32}$ such that each node in the network receives the same input from the inter-layer coupling strength.
\begin{figure}
    \centering
    \includegraphics[width=0.5\linewidth]{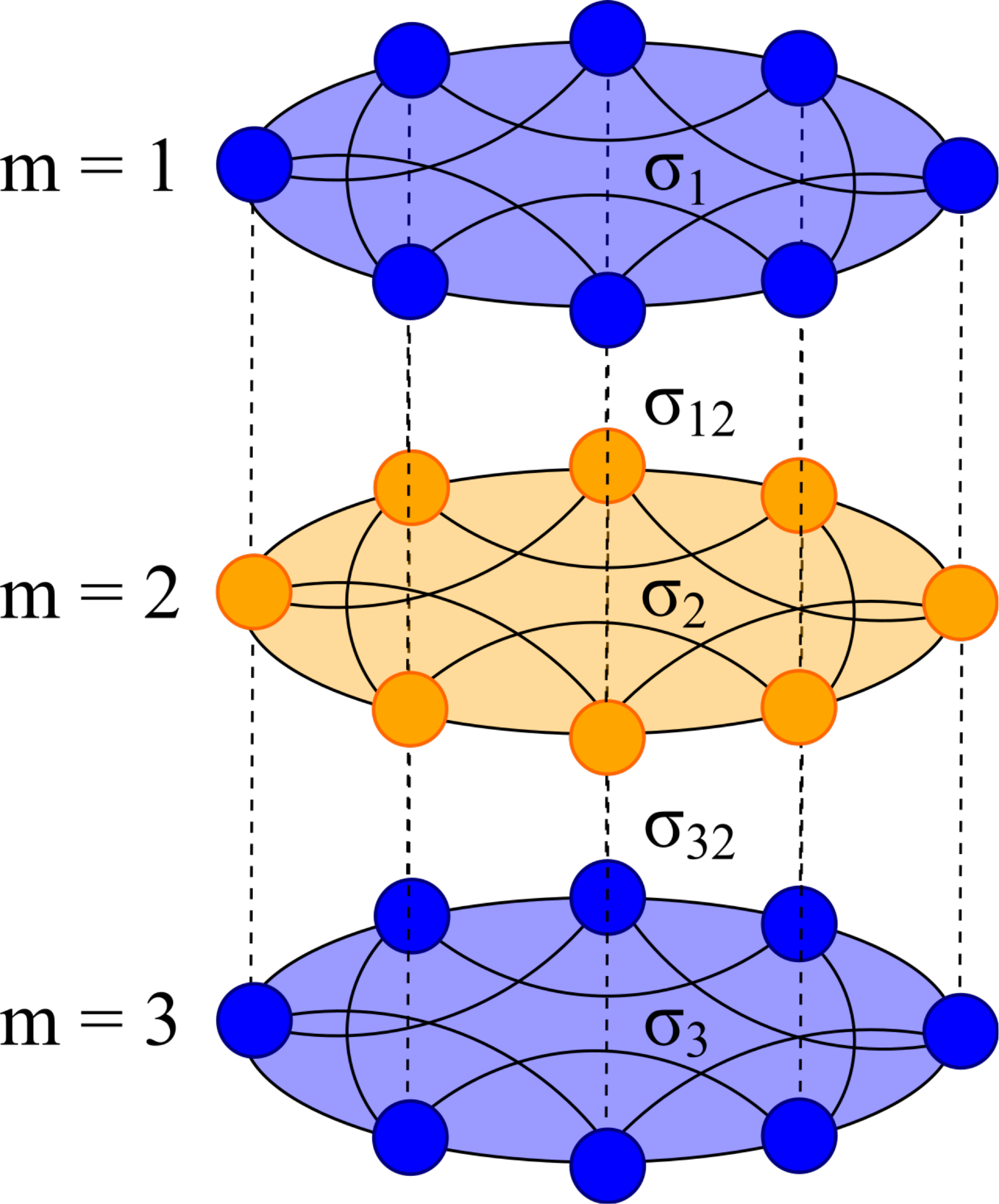}
    \caption{Schematic representation of the three-layer multiplex network: each layer is formed by $N$ non-locally coupled elements. The outer layers $m = 1$ and $m = 3$ are indirectly connected via the middle layer. The intra-layer and inter-layer couplings are denoted by $\sigma_{m}$ and $\sigma_{ml}$, respectively, where $\sigma_{12} = \sigma_{32}$.}
    \label{fig:3dlayer}
\end{figure}
In Eq.~(\ref{eq:multiplex}), the first term defines the dynamics of each individual node, the second term gives the nonlocal intra-layer coupling, and the third term corresponds to the inter-layer coupling, as illustrated in Figure\,\ref{fig:3dlayer}. In the present paper we will focus on a multiplex setup, namely identical topology within each layer, and pairwise coupling between the corresponding nodes from the neighbouring layers. 

In each individual layer, it is possible to observe chimera patterns depending on the range of nonlocal coupling and the strength of the couplings~\cite{OME11}. Note that chimera states in networks of time-discrete coupled maps differ from the chimera states in networks of time-continuous oscillators. Chimera patterns, consisting of coherent and incoherent domains, emerge as a result of the break-up of the smooth wavelike profiles~\cite{OME11,OME12,HAG12}, with an even number of coherent domains, and successive coherent domains appear in anti-phase to one another, while the incoherent domains are characterized by spatial chaos. In the following, we will analyze the synchronization of such complex patterns between the layers of our three-layer network.

%******************************

\section{\label{sec:Results}Numerical Results}

\subsection{Relay synchronization}

To measure the synchronization between two layers $m,l$ the global inter-layer synchronization error $E^{ml}$ for time-discrete systems will be used:
\begin{equation}
\label{eq:syncerrtotal}
    E^{ml} = \lim_{T \rightarrow \infty} \frac{1}{NT}  \sum^T_{t=0} \sum^{N}_{i=1} \bigl|\bigl|z^l_i(t) - z^m_i(t)\bigr|\bigr|,
\end{equation}
where $||\cdot||$ is the Euclidean norm and $T$ is the number of time-steps. For relay synchronization we need to measure the synchronization error between the first and the third layer $E^{13}$ and between the first and the second layer $E^{12}$. Thus, relay synchronization appears when $E^{13} = 0$ and $E^{12} \neq 0$. The synchronization error $E^{ml}$ is normalized by the total amount of nodes $N$ of one layer for better comparison of systems with different size $N$~\cite{SAW18c}.
\\
To characterize the synchronization of chimera patterns between the layers in more detail, we define the local synchronization error in dependence of each node $i$: 
\begin{equation}
\label{eq:syncerrnodes}
    E^{ml}_i = \lim_{T \rightarrow \infty} \frac{1}{T}  \sum^T_{t=0} \bigl|\bigl|z^l_i(t) - z^m_i(t)\bigr|\bigr|,
\end{equation}

\begin{figure}[t!]
    \centering
    \includegraphics[width=1\linewidth]{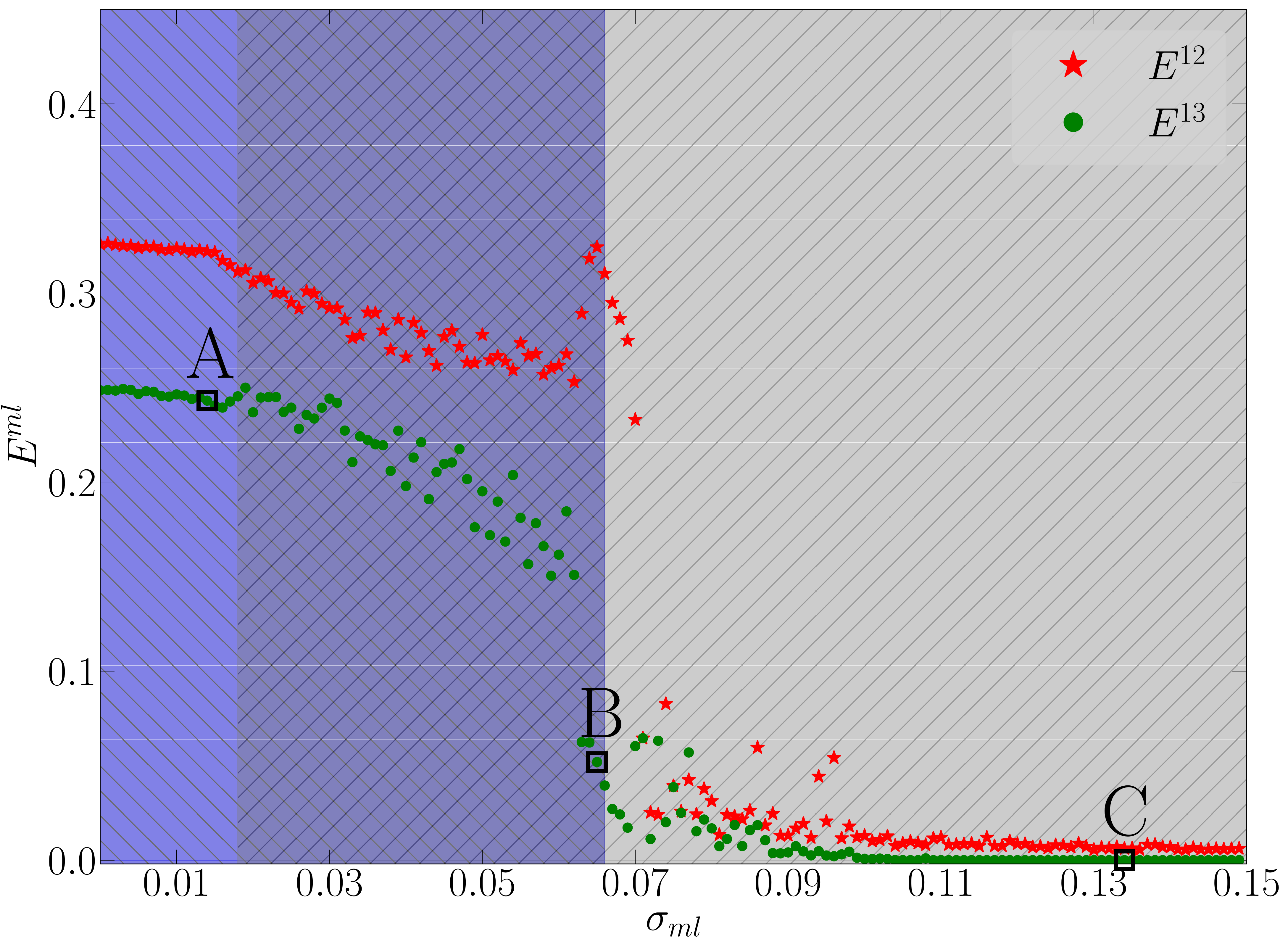}
    \caption{Global synchronization error between the first and second layer $E^{12}$ (yellow stars) and between the first and third layer $E^{13}$ (blue dots), respectively, versus $\sigma_{ml}$. For small inter-layer coupling strength the system is completely desynchronized, and with increasing $\sigma_{ml}$ the overall synchronization increases. The regimes of phase chimeras and amplitude chimeras are shown by blue and gray shading, respectively. Note that there is an overlap between the two. The black squares A (left), B (middle) and C (right) mark three values of $\sigma_{ml}$ which are further analyzed in Fig.\,\ref{fig:rm28sm22row}. Other parameters: $\sigma_m = 0.22$, $r_m = 0.28$, $N_m = 1000$, $a = 3.8$. All simulations are run for $5000$ time steps and the synchronization error is averaged over the last $50$ steps to avoid transient effects.}
    \label{fig:rm28sm22main}
\end{figure}

\begin{figure}[t!]
    \centering
    \includegraphics[width=0.9\linewidth]{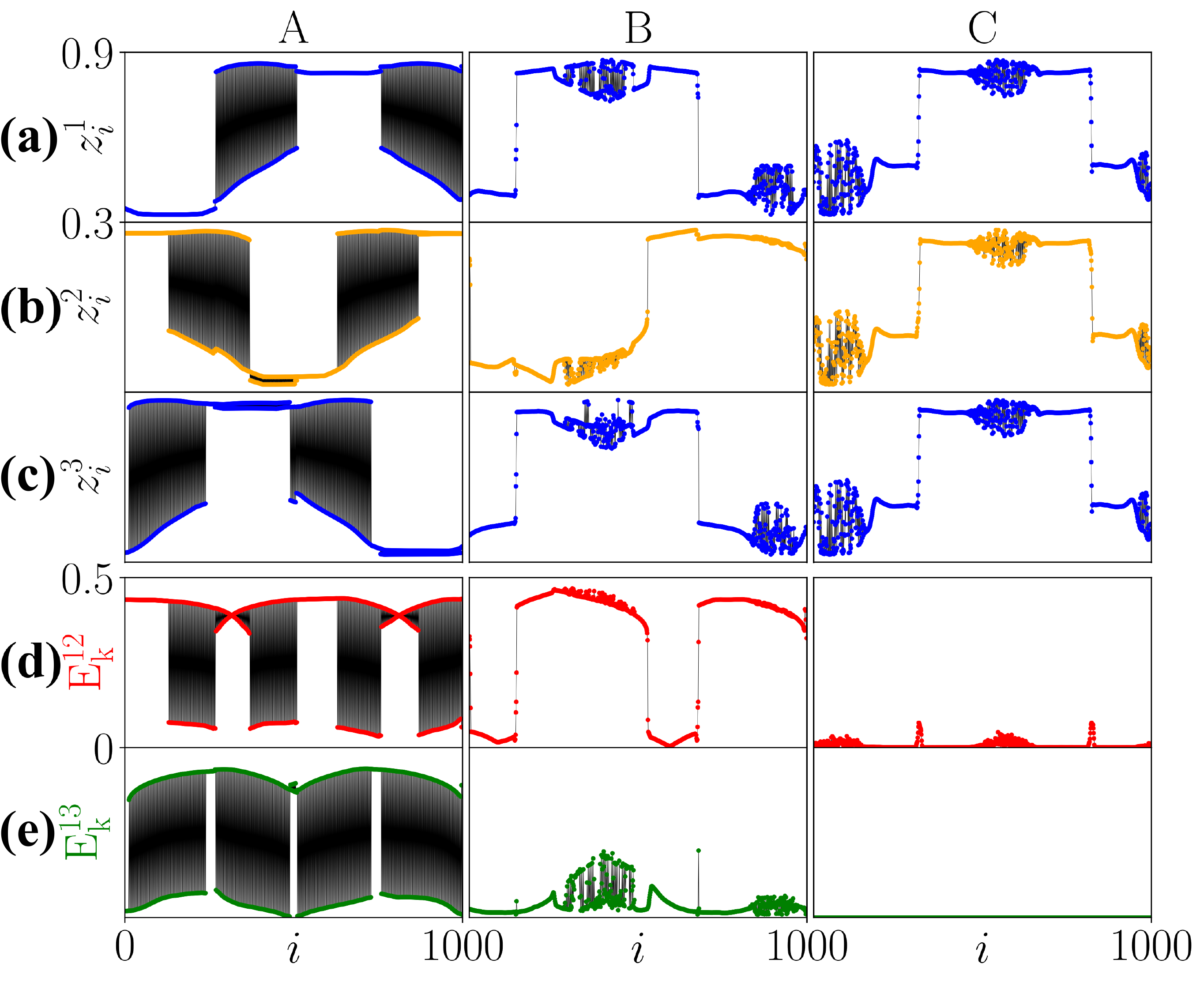}
    \caption{Transition to relay synchronization for three selected inter-layer coupling strengths $\sigma_{ml}$, marked by A (left panel), B (middle panel) and C (right panel) for $\sigma_{ml} = 0.015, 0.066, 0.135$ respectively (black squares in Fig.\,\ref{fig:rm28sm22main}). (a),(b),(c): snapshots of $z^{1}_i$, $z^{2}_i$, $z^{3}_i$, (d) local synchronization error between the first and second layer $E^{12}_k$, (e) $E^{13}_k$ versus node index $i$. Other parameters as in Fig.\,\ref{fig:rm28sm22main}.}
    \label{fig:rm28sm22row}
\end{figure}

Fig.~\ref{fig:rm28sm22main} presents simulations of the triplex network in the regime where the individual uncoupled layers exhibit chimera states. The plot shows the global synchronization error $E^{ml}$ (Eq.\,(\ref{eq:syncerrtotal})) versus the inter-layer coupling strength $\sigma_{ml}= \sigma_{12}=\sigma{32}$. The green dots represent the global synchronization error $E^{13}$ between the first and the third layer. The orange stars represent the synchronization error $E^{12}$ between the first and the second layer. 

The synchronization error between the first and third layer stays below the error between the first and second layer. This is an indicator of relay synchronization where the second layer plays the role of a relay layer. The overall synchronization of the considered 3-layer network tends to increase with increasing inter-layer coupling strength. At a certain value of $\sigma_{ml}$ the synchronization error drops sharply within a short range of $\sigma_{ml}$ to zero where $E^{13}$ reaches zero before $E^{12}$.

Exemplary snapshots of the dynamics for selected inter-layer coupling strength $\sigma_{ml}$ from regions of interest of Fig.\,\ref{fig:rm28sm22main} are shown in Fig.~\ref{fig:rm28sm22row}: (a), (b), (c) depict the dynamical variable $z^{1}_i$, $z^{2}_i$, $z^{3}_i$ of the first, second and third layer, respectively. (d), (e) represent the local synchronization error between the first and second layer $E^{12}_k$ and between the first and third layer $E^{13}_k$, respectively, for each node as introduced in Eq.\,(\ref{eq:syncerrnodes}). This is an adequate space-resolved measure for the synchronization of chimera patterns.

The first column A in Fig.~\ref{fig:rm28sm22row} corresponds to weak inter-layer coupling strength $\sigma_{ml}= 0.015$ and to a large value of $E^{13}$ marked by a black square in Fig.\,\ref{fig:rm28sm22main}. The low value of $\sigma_{ml}$ has almost no effect on the triplex network since no synchronization can be observed in the snapshots. Also the local synchronization errors $E^{12}_k$, $E^{13}_k$ in panels (d), (e) show high values.\\

The middle column B in Fig.~\ref{fig:rm28sm22row} is associated with an intermediate value of $\sigma_{ml}=.066$ in the transition zone from high to low global synchronization error $E^{13}$ marked by a black square in Fig.~\ref{fig:rm28sm22main}. In Fig.~\ref{fig:rm28sm22row} the snapshots of the dynamical variables in (a) and (c) show slightly different small-amplitude oscillation but spatial synchrony. The second layer (b) shows desynchronized behavior compared to the two outer layers, i.e., the coherent and incoherent domains are incongruous except for a small overlap, see (d). The stronger inter-layer coupling leads to adaptation of the location of incoherent and coherent domains on the two rings in the outer layers mediated by the relay layer. Nevertheless, the oscillations of each node are still not perfectly synchronized, which is shown by the synchronization error $E^{13}$ in panel (e): The coherent part of the outer layers are synchronized, whereas the incoherent ones are not. We call such a scenario {\it partial relay synchronization}.\\

The third column C in Fig.~\ref{fig:rm28sm22row} extracts further information about the behavior of the system for large inter-layer coupling strength $\sigma_{ml} = 0.135$ marked in Fig.~\ref{fig:rm28sm22main} by a black square. Here, full relay synchronization can be observed. The coherent and incoherent domains coincide in the first and third layer (panels (a),(c)). Also the second layer (b) behaves qualitatively in a similar way, however, comparing the local synchronization error, one can see that $E^{13}_k$ is exactly equal to zero for each node in the layer, whereas the local synchronization error $E^{12}_k \neq 0$ demonstrates that the second layer is not completely synchronized with its outer counterparts. It acts like a relay layer which stabilizes synchrony between the two outer layers, which is a signature of relay synchronization between the outer layers of the network.
\begin{figure}[t!]
    \centering
    \includegraphics[width=0.9\linewidth]{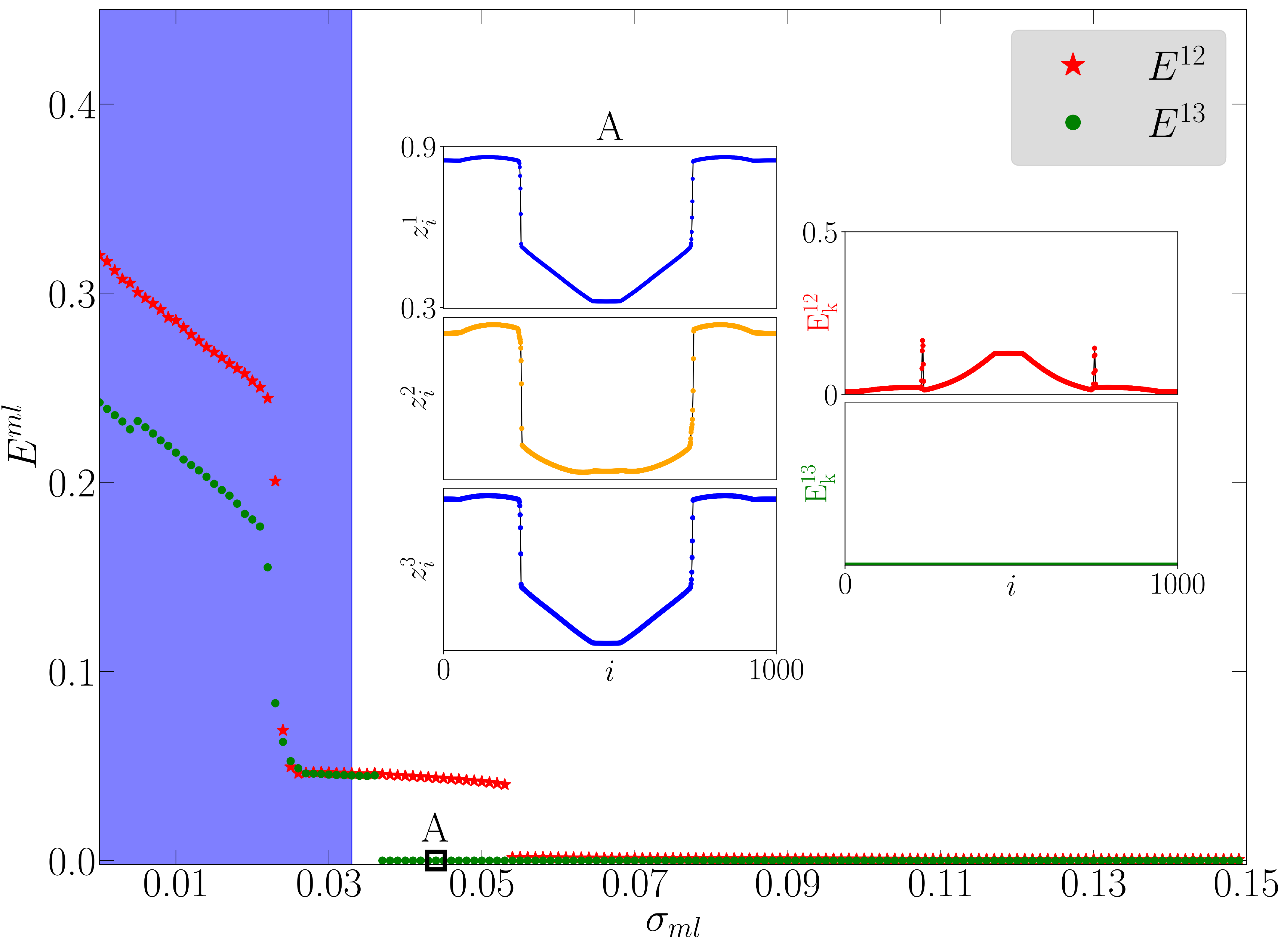}
    \caption{Same as Fig.~\label{fig:rm28sm22main} for $\sigma_m = 0.32$, $r_m = 0.30$. The inset shows the snapshots and local synchronization errors $E^{12}_k$, $E^{13}_k$ corresponding to the black square A in the main figure $\sigma_{ml} = 0.045$.}
    \label{fig:rm30sm32main}
\end{figure}

\subsection{Transition from phase to amplitude chimeras}

In continuous-time oscillator networks, two different types of chimera states have been observed and studied. The so-called {\it phase chimeras} demonstrate coexistence of oscillator groups with synchronized and desynchronized phases. The {\it amplitude chimeras} are patterns where all oscillators are phase locked, but synchronized and desynchronized domains appear in terms of oscillator amplitudes~\cite{ZAK14}. In recent works on networks of time-discrete coupled maps, these two notions have been also used to define phase and amplitude chimeras in terms of incoherent domains with large (phase flips) and small variations, respectively~\cite{BOG16,BOG16a}. In this work, we will use this terminology as well.
 
The regimes of phase chimeras and amplitude chimeras are shown in Fig.~\ref{fig:rm28sm22main} by blue and gray shading, respectively. Note that there is an overlap between the two. Fig.~\ref{fig:rm28sm22row} demonstrates three examples of patterns in each of three layers for small, intermediate and large inter-layer coupling strength, marked as $A,B,C$ in Fig.~\ref{fig:rm28sm22main}. For weak inter-layer coupling strength (A), we observe phase chimera states in each layer of the multiplex network. The first column in Fig.\,\ref{fig:rm28sm22row} shows in black the incoherent domains in all three layers where we find an irregular sequence of values on the upper and lower coherent branch, corresponding to phase flips of $\pi$. Pronounced phase chimera states like these cannot be observed for larger values of $\sigma_{ml}$. 

%Referring to the analytic approximation in in Sec.\,\ref{sec:AnalyticalResults} the approximate condition for phase chimeras $\sigma_m + 2 \sigma_{ml} < 0.44$ (Eq.\,(\ref{eq:finalfinal})) is only fulfilled for small $\sigma_{ml}$ by considering anti-phase oscillation between layers $1$ and $2$. Indeed, for the left column $\sigma_m + 2 \sigma_{ml} = 0.25$, and for the right column $\sigma_m + 2 \sigma_{ml} = 0.49$.
%The middle column corresponds to the transition zone with $\sigma_m + 2 \sigma_{ml} = 0.35$ which is within the possible error of our approximation. 
%The break of symmetry which entered through the considered 3-layer-multiplex network where each layer represents a symmetric ring could be an explanation for the difficulty of inducing similar strong chimera states as in the one-layer case. 
With increasing inter-layer coupling strength $\sigma_{ml}$, amplitude chimera states are born within the upper and lower coherent branch (B). They are characterized by small variations in the incoherent domains and arise since each layer consists of a closed-ring of time-discrete oscillators which are additionally coupled to an extra node from another layer. These amplitude chimeras can be perfectly synchronized in the considered network between the outer layers as indicated by the local synchronization error $E^{13}_k = 0$ in the right column (C) of Fig.~\ref{fig:rm28sm22row} (e). On the other hand ,$E^{12}_k \neq 0$ proves that the middle layer is not fully synchronized, i.e., relay synchronization arises. The corresponding plots of $E^{12}$ and $E^{13}$ in Fig.\,\ref{fig:rm28sm22row} illustrate that the necessary condition for phase chimeras derived in Sec.\,\ref{sec:AnalyticalResults} is violated. 

The intra-layer parameters play also an important role in the synchronization scenarios for the whole network. Fig.~\ref{fig:rm30sm32main} shows the dependence of the global synchronization error on the inter-layer coupling strength in the case of larger coupling range $r_m=0.3$ inside the layers and stronger intra-layer coupling strength $\sigma_m=0.32$. In this case, only very weak inter-layer coupling results in the observation of phase chimeras in all layers, and its further increase moves the system towards the formation of smooth profiles. However, there is indeed an intermediate region where distinct relay synchronization between the outer layers occurs, bounded by two abrupt  discontinuous transitions of $E^{13}$ and $E^{12}$ to zero, and all this occurs for smooth profiles in contrast to Fig.~\ref{fig:rm28sm22main}. The insets in Fig.\,\ref{fig:rm30sm32main} depict the corresponding snapshots and local synchronization errors at point A.

\section{\label{sec:AnalyticalResults}Analytical Results}

Using a similar argument as in \cite{OME12,HAG12,SEM15a} for single-layer networks, we can calculate analytically the critical intra-layer coupling strength $\sigma_c$ for the onset of phase chimeras in case of a triplex network. The coherent solutions $z_i^m(t)$ approach a smooth profile $z(x^m,t)$ in the continuum limit $N \rightarrow \infty$ which leads to one equation per layer, where $x^m$ denotes the space coordinate in layer $m$.  Denoting the intra-layer coupling strength $\sigma_1 = \sigma_2 = \sigma_3 = \sigma$, the inter-layer coupling strength $\sigma_{12} = \sigma_{32} = \Phi$, and the coupling range $R_1 = R_2 = R_3 = R,$ hence $r_m = \frac{R_m}{N} = r$, we obtain the following equations for each layer:
        \begin{eqnarray*}
        \label{eq:1}
            z^{(1)}(x,t+1) &= f(z^{(1)}(x,t)) \\ 
            &+\frac{\sigma}{2r} \int_{x-r}^{x+r} [f(z^{(1)}(y,t))- f(z^{(1)}(x,t))]dy \\
            &+\Phi~[f(z^{(2)}(x,t))- f(z^{(1)}(x,t))]
        \end{eqnarray*}
        
        \begin{eqnarray*}
        \label{eq:2}
            z^{(2)}(x,t+1) &= f(z^{(2)}(x,t)) \\
            &+\frac{\sigma}{2r} \int_{x-r}^{x+r} [f(z^{(2)}(y,t))- f(z^{(2)}(x,t))]dy \\
            &+ \frac{\Phi}{2}~[f(z^{(1)}(x,t))- f(z^{(2)}(x,t))]\\
            &+ \frac{\Phi}{2}~[f(z^{(3)}(x,t))- f(z^{(2)}(x,t))] 
        \end{eqnarray*}
       
        \begin{eqnarray*}
        \label{eq:3}
            z^{(3)}(x,t+1) &= f(z^{(3)}(x,t)) \\
            &+\frac{\sigma}{2r} \int_{x-r}^{x+r} [f(z^{(3)}(y,t))- f(z^{(3)}(x,t))]dy \\
            &+\Phi~[f(z^{(2)}(x,t))- f(z^{(3)}(x,t))] 
        \end{eqnarray*}
   
To derive a relation for the critical intra-layer coupling strength $\sigma_c$ we can conduct the following steps. First, we transform the equation for layer 1 into:
         \begin{eqnarray*}
         \label{eq:last}
            z^{(1)}(x,t+1) &= (1-\sigma-\Phi)f(z^{(1)}(x,t)) \\
            &+\frac{\sigma}{2r} \int_{x-r}^{x+r} f(z^{(1)}(y,t))dy \\
            &+ \Phi f(z^{(2)}(x,t)) 
        \end{eqnarray*}
Consider a solution of Eq.\,(\ref{eq:last}) in the form of a smooth wave profile with wave number $k=1$, i.e., with wavelength $\lambda$ equal to the length of the ring $L$, and period-2 dynamics for each layer.  We can reduce the dynamics to even and odd time steps $z_0(x)$ and $z_1(x)$, respectively. This yields the following spatial derivatives for $j = 0,1$ where we distinguish between two cases where in the first case the two layers 1 and 2 are \textit{in-phase} and in the second case they are \textit{anti-phase}:
        \begin{eqnarray*}
        \label{eq:derivation}
            z'_{1-j}(x) &= (1-\sigma-\Phi)~f'(z_j(x))z'_j(x) \\
            &+\frac{\sigma}{2r}[f(z_j(x+r))-f(z_j(x-r))]\\ 
            &+ \Phi \begin{cases} f'(z_j(x))~~z'_j(x)~~~ \text{in-phase}\\
                                f'(z_{1-j}(x))~z'_{1-j}(x)~~~ \text{anti-phase}
                    \end{cases} 
        \end{eqnarray*}
At the point $x^*$ where the smooth profile breaks up and chimera states are born, the spatial derivative becomes infinite. Considering that $z'_0(x^*), z'_1(x^*)$ diverge to infinity, we neglect the term without derivative $\frac{\sigma}{2r}[f(z_j(x+r))-f(z_j(x-r))]$. Therefore we obtain for the odd and even time steps:
    \begin{eqnarray*}
    j=0:~~~ z'&_1(x^*) = (1-\sigma-\Phi)~f'(z_0(x^*))~z'_0(x^*)\\
       &+ \Phi \begin{cases} f'(z_0(x^*))~z'_0(x^*)~~~ \text{in-phase}\\
        f'(z_{1}(x^*))~z'_{1}(x^*)~~~ \text{anti-phase}
                    \end{cases}\\
                    \\
       j=1:~~~ z'&_0(x^*) = (1-\sigma-\Phi)~f'(z_1(x^*))~z'_1(x^*)\\
       &+ \Phi \begin{cases} f'(z_1(x^*))~z'_1(x^*)~~~ \text{in-phase}\\
        f'(z_0(x^*))~z'_0(x^*)~~~ \text{anti-phase}
                    \end{cases}\\
    \end{eqnarray*}
Since we choose the parameters for every layer in the regime of wave number $k=1$ and period-2 dynamics in time, we assume that the spatial derivatives at even and odd time steps satisfy $z'_0(x) = -z'_1(x)$ and that at the break-up point $z_0(x^*) = z_1(x^*)  \equiv z^*$ and $z'_0(x^*) \equiv z'^*$ holds.
Multiplying the upper equations (in-phase) for $j=0$ and $j=1$ we obtain for the two cases:\\
(i) First and second layer in-phase:
    \begin{eqnarray*}
    \label{eq:phase}
             (z'^*)^2 &= (1-\sigma-\Phi)^2~f'(z^*)^2~(z'^*)^2\\
                    &~~~+ 2 \Phi~(1-\sigma-\Phi)~f'(z^*)^2~(z'^*)^2\\
                    &~~~+ \Phi^2~f'(z^*)^2~(z'^*)^2 \\
    \end{eqnarray*}
		which yields
		\begin{equation}\label{eq:crit_inphase}
        1 = (1-\sigma)^2~f'(z^*)^2 
    \end{equation}
\\
(ii) First and second layer anti-phase:
    \begin{eqnarray*}
    \label{eq:antiphase}
             (z'^*)^2 &= (1-\sigma-\Phi)^2~f'(z^*)^2~(z'^*)^2\\
                    &~~~- 2 \Phi~(1-\sigma-\Phi)~f'(z^*)^2~(z'^*)^2\\
                    &~~~+ \Phi^2~f'(z^*)^2~(z'^*)^2 \\
    \end{eqnarray*}
		which yields
		\begin{equation}\label{eq:crit_antiphase}
        1 = (1-\sigma-2\Phi)^2~f'(z^*)^2
    \end{equation}
To derive an approximation for the critical coupling strength where the smooth profile breaks up, let $z^*$ be the fixed point of the local logistic map: $z^* = f(z^*) = az^*(1-z^*)$, hence $z^* =1 - 1/a \approx 0.737$ with $a=3.8$. We solve the equations (\ref{eq:crit_inphase}) and (\ref{eq:crit_antiphase}) with $f'(z^*) = a(1-2z^*)=2-a$:\\
(i) in-phase:
    \begin{equation}
    \label{final:phase}
        1-\sigma = \pm \frac{1}{|f'(z^*)|} 
    \end{equation}
(ii) anti-phase:
    \begin{equation}
    \label{eq:finalantiphase}
        1-\sigma-2\Phi = \pm \frac{1}{|f'(z^*)|} 
    \end{equation}
    \\
We choose the plus sign of Eqs. (\ref{final:phase}) and (\ref{eq:finalantiphase}) since the lower value of $\sigma$ represents the threshold where the smooth profile breaks up with decreasing coupling strength, and with $|f'(z^*)| =  1.8$, we derive the condition for the onset of phase chimeras
\begin{eqnarray}
\label{eq:finalfinal}
    \left.  
    \begin{array}{@{}lr@{}}
    %\multirow{}{}
    ~\text{if layers 1 and 2 are in-phase:}~ &\sigma \nonumber \\
    ~\text{if layers 1 and 2 are anti-phase:}~~~ \sigma~ + &2\Phi \nonumber \\
    \end{array}\right \} &\approx 0.44
\end{eqnarray}

Thus phase chimeras only exist if the effective coupling strength is below this critical value, i.e., $\sigma<0.44$ if layers 1 and 2 oscillate in-phase, and $\sigma + 2\Phi <0.44$ if the two layers oscillate anti-phase.
Interestingly, for in-phase oscillation the inter-layer coupling $\Phi$ has no effect on the critical coupling strength. On the other hand, if we consider anti-phase oscillations, the inter-layer coupling strength $\Phi$ has to be taken into account. This explains why phase chimeras can exist in single layers, but with increasing inter-layer coupling strength they disappear. Of course the condition $\sigma + 2\Phi <0.44$ was derived under very crude approximations and therefore only a rough estimate like $\sigma + 2\sigma_{ml} <0.4$ can be applied.  Indeed, in Figs.~(\ref{fig:rm28sm22main}) and (\ref{fig:rm28sm22row}) we find phase chimeras only for small $\sigma_{ml}$, as in the marked point A of Fig.~(\ref{fig:rm28sm22main}) where $\sigma_m + 2\sigma_{ml} = 0.25$. In point C, i.e., the right column of Fig.\,\ref{fig:rm28sm22row} with $\sigma_m + 2\sigma_{ml} = 0.49$ we cannot observe phase chimeras anymore, and point B, i.e., the middle column with $\sigma_m + 2\sigma_{ml} = 0.35$ seems to correspond to a transition zone of both regimes within the possible error of our approximation.

\section{\label{sec:Conclusion}Conclusion}

We have demonstrated that multiplex networks of time-discrete maps allow for intriguing relay synchronization scenarios, where distant layers synchronize in spite of the absence of direct connections between them. We have analyzed relay synchronization in a three-layer network of logistic maps, with nonlocal coupling topologies within the layers. The uncoupled nodes are characterized by chaotic dynamics, however, due to the coupling the logistic map gives rise to spatially smooth profiles or chimera patterns, and can perform periodic dynamics in time. We observe two types of chimera states with large (phase chimeras) and small (amplitude chimeras) variations of the dynamical variable, and demonstrate the transition from one type of chimera (phase chimeras) to the other with increasing inter-layer coupling strength. We find regimes of relay synchronization between amplitude chimeras in the outer layers, as well as partial relay synchronization of chimera states in the two outer layers in the form of intriguing double chimeras, where the coherent domains in both layers are synchronized, while the incoherent ones are not. By choosing an appropriate value for the inter-layer coupling strength we can switch between the different synchronization scenarios.\\

We have provided an analytic approximation for the coupling strengths of the network necessary for phase chimeras, and have explained with this our observation that phase chimeras disappear with increasing inter-layer coupling strength.
Our findings show that weak inter-layer coupling with small $\sigma_{ml}$ is crucial for relay and partial relay synchronization in the networks.
The advantage of this simple paradigmatic model is that it allows for analytical insight into the dynamics of patterns. 
Our results can be useful for the analysis of relay synchronization in multiplex and multilayer networks with more complex dynamics of the individual nodes, and for numerous applications where relay synchronization occurs, e.g., in neuronal systems \cite{SAW18c}.

\section{Acknowledgement}
Funded by the Deutsche Forschungsgemeinschaft (DFG, German Research Foundation)- Projektnummer 163436311- SFB~910.

\bibliography{ref}  
\bibliographystyle{prsty-fullauthor}
%\bibliographystyle{eplbib}
%\bibliographystyle{prwithtitle}  
%\begin{thebibliography}{10}
%\end{thebibliography}

\end{document}